# Energy-dependent scaling of incoherent spectral weight and the origin of the waterfalls in high-$T_c$ cuprates

Qiang Wang [1], Zhe Sun [1], Eli Rotenberg [2], Helmuth Berger [3], Hiroshi Eisaki [4], Yoshihiro Aiura [4] and D. S. Dessau [1]

[1.] Department of Physics, University of Colorado, Boulder, CO 80309-0390, USA

[2.] Advanced Light Source, Lawrence Berkeley National Laboratory, Berkeley, California 94720, USA

[3.] Institut de Physique de la Matière Complexe, EPFL, 1015 Lausanne, Switzerland

[4.] AIST Tsukuba Central 2, 1-1-1 Umezono, Tsukuba, Ibaraki 305-8568, Japan

**The exotic physics in condensed matter systems, such as high-$T_c$ superconductivity in cuprates[1], is due to the properties of the elementary excitations and their interactions. The dispersion of the electronic states revealed by angle-resolved photoemission spectroscopy (ARPES) [2, 3] provides a chance to understand these excitations. Recently, a "high energy anomaly" or "waterfall-like" feature in cuprates' dispersion has been reported [4, 5, 6, 7, 8, 9, 10, 11, 12] and studied theoreticaly[13, 14, 15, 16, 17]. Most of the current views argue that it is the result of some many-body effect at a specific high energy scale (e.g. ~ 0.3eV), though there are other arguments that this is an artificial effect[10, 11, 12]. Here, we report a systematic ARPES study on the "high energy anomaly" in Bi2212 samples over multiple Brillouin zones and with a large variety of ARPES matrix elements. We find that the incoherent weight of the electron spectral function at high binding energy is intimately linked to the energy of the dispersive coherent weight through an unexpected but simple relationship with no special energy scales. This behavior in concert with strong k-dependent matrix element effects gives rise to the heavily studied "waterfall" behavior.**

Fig. 1(a) shows an unsymmetrized and unfolded Fermi surface map of a Pb-Bi2212 sample over more than two Brillouin zones (BZ). In the plot, the Fermi surface contributions from the main band and the shadow band are clearly resolved. The superstructure or umklapp band is highly suppressed since Pb substitutes into the Bi-O plane and removes the superstructure replica[18]. The intensity maps along cuts C1 and C2 are shown in panels (b) and (c), respectively. These two cuts are along equivalent directions in momentum space and so we should expect to see the same dispersion. In fact, they appear quite different, which is an artifact of the photoemission matrix elements. In fig. 1(b) the dispersion starts vertically diving from a Binding Energy (BE) ~0.4eV and there is no band minimum observed down to BE=1.1eV. In fig. 1(c) on the other hand, a band minimum at BE~0.5eV is clearly seen. Additionally, in fig. 1(d), which is the intensity map at BE=0.6eV, the spectral intensity is highly suppressed along the two zone diagonal directions in the first zone. In the second zone however, the intensity is only suppressed along the horizontal direction. Such a suppression pattern is likely a result of the $d_{x^2-y^2}$ orbital symmetry of the Cu states being probed[19]. The spectrum at cut C3 ($k_y$=0.25π/a) in fig. 1(d) is shown in fig. 1(e). In this spectrum, the dispersion breaks at $k_x$=±0.25π/a as indicated by the arrows and this broken position just matches the intensity suppression pattern shown in fig. 1(d). Further, the two low intensity stripes in panel (d) at $k_x$=±0.25π/a mean that this intensity suppression effect persists throughout the whole binding energy range down to 1.1eV.

In ARPES, the measured spectral intensity can be expressed as *I*(**k**, *ω*)=*I$_0$*(**k**, *υ*, **A**)*f(ω)A(***k**, *ω)*, where *I$_0$*(**k**, *υ*, **A**) is proportional to the one-electron matrix element, *f(ω)* is the Fermi function and *A(***k**, *ω)* is the single-particle spectral function[3]. The single-particle spectral function should have the same value at all equivalent points in different BZs and the Fermi function does not contain any momentum information, so one must take the matrix element effect as the reason for the difference of the spectra in the first and second zones. Also the matrix element effect may explain why the apparent dispersion at high energy scales is almost perfectly vertical, which in this picture would indicate it is not a real dispersion.

More experimental data sets are shown in fig. 2 panels (a)-(d), with the locations of the cuts shown in the inset of panel (c). The dotted black lines show the dispersion obtained from Momentum Distribution Curve (MDC) fits of the data, while the red dashed lines show the dispersion obtained from tight-binding fits to the data, as will be discussed later and in the appendix. The spectra taken away from the diagonal matrix element minima of fig. 1(d) show that the spectral peaks of the data match the tight-binding dispersion (panels (b) and (c)) while others, which show the waterfall feature most clearly (a and d), are not close. However, whether or not the matrix element effect is present, all cuts show strong spectral weight beyond the band minimum. To reveal the properties of this high energy spectral weight, four Energy Distribution Curves (EDCs 1 to 4) with peak positions at -0.03eV, -0.1eV, -0.2eV and -0.3eV are chosen from the left side band in fig. 2(d). Another four EDCs (1' to 4') are chosen from the right side band in the same manner. EDC0 is chosen from the momentum space where no dispersive band is present ($k>k_F$). This spectrum, perhaps due to elastic scattering of the electrons from the occupied ($k<k_F$) k points, will be helpful for removing a background term from the other spectra.

The raw spectra EDC0 and EDC1' to EDC4' are plotted in fig. 3(a). Panel (b) shows the same spectra except that they have all had the "background" spectrum EDC0 subtracted. Comparing fig. 3(a) and fig. 3(b), we find that: first, in both panels all the EDCs have a strong tail at high binding energy which is consistent with the strong spectral weight below the band minimum as seen in fig. 2; second, all the EDCs in fig. 3(a) have a rising tail at high binding energy range which is not shown in fig. 3(b) after subtracting EDC0; finally, there is a clear trend that as the EDC peak position gets deeper, the tail gets higher.

We also subtract EDC0 from the weak-side spectra EDC1 to EDC4 and overlay the background subtracted EDCs from both sides of the band in the same plot to compare their lineshapes. The results are plotted in figs 3(c) to 3(f) and the EDCs from the weak-side are rescaled with different normalization factors (listed in the figures) to match the peak intensities of those from the strong-side. The difference in those renormalization factors is also a sign of the k-dependent matrix element effect. It is clearly found that the spectra from both sides have the same lineshape once the spectrum EDC0 has been subtracted off in spite of the greatly varying intensity. This is convincing evidence that the empirical background subtraction method we found here returns spectra which are a very close approximation to the true intrinsic lineshapes.

To get a further understanding of the behavior of the spectral weight at high binding energy, we adopt a method originated by Shirley, which was originally developed to distinguish the background of inelastic "secondary" photoelectrons from un-scattered "primary" photoelectrons[20]. In this method the background is separated from the spectrum by the formula

$$P(E) = R(E) - B(E) = R(E) - \kappa \int_{E}^{0} P(E')dE' \qquad (1)$$

where $R(E)$ is the raw EDC, $B(E)$ is the background term and $P(E)$ is the primary or peak of the EDC after the Shirley background correction. In the original treatment by Shirley, the proportionality factor $\kappa$ indicated the fraction of primary photoelectrons that underwent scattering events to appear as lower kinetic energy secondary photoelectrons. In our case, it tells us the relative strengths of the peak and background weights, as well as just giving us a convenient mechanism to separate out the peak from the background, (see fig. 4(b), which are three EDCs taken from the data of fig. 4(a) at $k_x$=0.06π/a, 0.15π/a and 0.24π/a). We found this method to be robust even in the instance of kink-induced "peak-dip-hump" structures in EDCs[21, 22] (see for example the middle figure in panel (b)). Fig. 4(c) shows the "Shirley background subtracted" result of panel (a). After this procedure, the spectrum shows a very clear band without the strong spectral weight at high binding energy and the dispersion matches the tight-binding fitting result very well.

With this method we analyzed spectra taken from many different places in momentum space throughout both the first and second BZs (see inset of panel 4(d)) taken under many different experimental conditions (photon energies, polarizations, etc.). A compilation of all these data is shown in fig. 4(d), which shows the $\kappa$ factors returned from each of the individual fit, plotted against the EDC peak positions obtained from the same fit. Fig. 4(d) indicates an unexpected but quite simple scaling relationship - the extracted $\kappa$ factor scales closely with the extracted EDC peak position, independent of experimental condition or location in momentum space. This means that there is a general relationship between the strength of the high energy tail and the EDC peak position. As the EDC peak gets deeper, the κ factor keeps rising, i.e. more spectral weight is transferred from the coherent part to the incoherent part as the coherent part goes to higher binding energy. To quantify this relation, we have parameterized the κ values as a function of energy, finding that a simple quadratic function does an adequate job, as shown in fig. 4(d). Additionally, there is no evidence of any breaks or steps at for example the often-cited anomalous energy scale of 300~500meV[5, 6, 7, 8, 9] in the "κ vs. EDC peak position" curve. This is further, more robust, evidence against any new high binding energy scale where the high energy anomaly begins.

In our fitting, the κ factor for the EDC with different peak position varies by almost an order of magnitude. This is strong evidence that the origin of the background removed here is NOT the inelastically scattered electrons originally considered by Shirley, because in that case the κ factor is expected to be essentially independent of the binding energy of the initial excited state. Additionally, even the smallest κ factor we extracted is still more than an order of magnitude larger than the κ factor for inelastically scattered electrons obtained from a valence band result on Bi2212[23]. We also note this analysis is fully consistent with our recently reported laser-excited ARPES data[24] with photon energy down to 6eV which show a much sharper quasi-particle excitation and a much weaker background. As shown in Fig. 4(e), the κ value extracted from these data is similar to that obtained from the synchrotron data, though the laser data are not available to deep binding energies so there are some more uncertainties in the determination of the background parameters.

Our analysis indicates that the deep spectral weight which makes up the "waterfall" structures is an intrinsic part of the spectral lineshape. This is fully in line with the consistent spectral lineshapes shown in panel (c)-(f) of fig. 3 which show that the tails and peaks of the EDCs have essentially identical k-dependent matrix element effects, i.e. they are from the same state with the same symmetry. This also is completely consistent with our observation that the E vs. k dispersion of the waterfall is always completely "vertical", i.e. in the E direction as opposed to the continuously varying E vs. k of a dispersing quasiparticle.

Our result differs from all earlier interpretations which ascribe the waterfalls as due to the coupling to a mode. One reason is the lack of any energy scale other than the bottom of the tight binding dispersion (~ 0.55eV), while another is the truly "vertical" dispersion, which is at odds with the theoretical models of coupling to a high energy mode[7, 15, 17]. And while some groups have plotted a dispersive E vs. k relation for the waterfall portion itself[5, 6, 7, 8, 9] we believe this dispersion to be an artifact of the energy and k-dependent matrix elements which can give the impression of a slight amount of dispersion away from the vertical direction.

While we now understand that the waterfalls are nothing more than the high binding energy tails of the EDCs, we have not yet come to an understanding of the origin of these tails. This has in fact been a very deep question in the field for almost two decades, dating back to Anderson's and others' attempts to explain the anomalous ARPES lineshape of the cuprates[25, 26] and continuing up to the present year in which the latest nodal laser-ARPES data were analyzed in terms of a Gutzwiller-projected non-Fermi Liquid lineshape[27]. This Gutzwiller lineshape is an example of a lineshape which intrinsically has a significant amount of its spectral weight in the high binding energy tail while the Fermi liquid[28] or the marginal Fermi liquid[29] lineshapes require a relative weak tail at high binding energy region. A similar "Shirley background" analysis done on the Gutzwiller lineshape using previously determined parameters and from fits performed only on nodal data and then over a much smaller energy range[27], returns κ factors having the similar "κ vs. EDC peak position" relation (not shown) which is consistent with our result, although the κ values are not quite the same as those extracted directly from the data.

## Figure legends:

Fig. 1. (Color). Fermi surface and matrix element effects. (a) An unsymmetrized and unfolded Fermi surface map of a Pb-doped Bi-2212 sample. Black dashed lines show the zone boundaries of the first and second Brillouin zones, with the first zone centered at (0, 0) and labeled by Γ. The Fermi surface contributions from the main band and the shadow band are highlighted by the black and red dashed circles, respectively. Red dashed lines (C1 and C2) show the directions which the spectra in panels (b) and (c) are taken. (d) ARPES intensity map at 600 meV below the Fermi energy. Thin black dashed lines indicate the "spectral weight suppression direction" by the matrix element effect. Red dashed line (C3) shows the direction which spectrum (e) is taken. All the data were taken at T=52K (superconducting state) using 90eV photons.

Fig. 2. (Color). Individual spectra and matrix element effects. (a)-(d) ARPES spectra taken along cuts (a), (b), (c) and (d) shown in the Fermi surface inset in panel (c). Red dashed lines represent the 6 parameter tight-binding model dispersion along each cut. The data points chosen for the tight-binding fitting are also indicated in the inset in panel (c) by the red-yellow circles. In plot (a), the thin black dashed lines represent the MDC peak dispersion. In plot (d), the thin white dashed line represents the MDC peak dispersion of the strong right-side band; the thin black dashed line is the reflection of the white dashed line. Spectra (a)-(c) were taken at T=25K; (d) was taken at T=50K; all data were taken with 52eV photons.

Fig. 3. (Color). Removal of background from elastically scattered electrons. (a) EDC0 and EDC1'-EDC4' extracted from and indicated in Fig. 2(d). (b) "Background subtracted" EDCs by subtracting EDC0 from EDC1'-EDC4'. (c)-(f) EDC line shape comparison of the "background subtracted" EDCs from both sides of the band of Fig. 2(d). The EDCs from the weak band side are scaled up to match the EDCs from the strong band side for line shape comparison. The different renormalization factors are shown in each of the panels.

Fig. 4. (Color). Characterization of the strength of the high energy incoherent spectral weight. (a) Spectrum with elastic background (e.g. EDC0) subtracted, which is an expanded view of the cut from fig. 2(b). Red dashed line represents the tight-binding dispersion. (b) "Shirley-background-like" fitting on three selected EDCs 1, 2 and 3 indicated in panel (a). In panel (b2), two black arrows are used to indicate the kink-introduced "peak-dip-hump" structure. (c) "Shirley-like-background" subtracted spectrum from panel (a). (d) "Shirley-background-like" proportionality or κ-factor vs. EDC peak position. Red boxes, blue triangles and pink circles represent the fitting result for the EDCs of Pb-doped Bi2212 sample taken from ALS BL10 (52eV photons, s polarization, 25K to 50K), ALS BL7 (90eV photons, p polarization, 52K) and ALS BL12 (90eV photon, p polarization, 20K) respectively. The green triangles represent the fitting result for the EDCs of optimal Bi2212 sample taken from SSRL BL5-4 (7eV photon, p polarization, 10K). The black dashed line represents the 2nd order polynomial fitting of all the data points. The inset shows spectral cuts in momentum space where we chose the EDCs for fitting. All the cuts in Fig. 2 are included. (e) Two low temperature (20K) nodal EDCs peaked at 20meV from ALS BL10 on a Pb-doped Bi2212 sample and 6eV laser system on an optimal doped Bi2212 sample. Even though the laser data are sharper and appear to have a smaller background, the κ factor indicating the strength of the incoherent background is similar.

## Methods:

**Experiment description:**
High quality single crystals of $Pb_xBi_{2-x}Sr_2CaCu_2O_8$ (Pb-Bi2212) with a $T_c$~85K and optimally doped $Bi_2Sr_2CaCu_2O_{8+\delta}$ (Bi2212) with a $T_c$~91K were prepared for the ARPES experiments, with the crystals cleaved in the UHV environments of the ARPES spectrometers. The Pb-Bi2212 samples were studied using Beamline 7.0.1 (BL7), Beamline 10.0.1 (BL10) and Beamline 12.0.1 (BL12) at the Advanced Light Source (ALS), Berkeley. The angular resolution of the experiments was approximately 0.3° and the energy resolution was 20-35 meV (depending upon photon energy). The optimally doped Bi2212 samples were studied using Beamline 5-4 (BL5-4) at Stanford Synchrotron Radiation Lightsource (SSRL) and with the 6eV laser ARPES system at the University of Colorado at Boulder. The angular resolution of the experiments was approximately $0.3°$ and the energy resolution was better than 5meV.

**Tight-binding fitting procedure:**

A six-parameter tight-binding model is used to fit the dispersion:

$$\varepsilon(k_x, k_y) = \mu - 2t(\cos ak_x + \cos ak_y) - 4t'\cos ak_x \cos ak_y - 2t''(\cos 2ak_x + \cos 2ak_y)$$
$$- 4t^{(3)}(\cos ak_x \cos 2ak_y + \cos ak_y \cos 2ak_x) - 4t^{(4)} \cos 2ak_x \cos 2ak_y + (...)$$

Six data points chosen for fitting are represented by the red-yellow circles in the inset of fig. 2(c). Notice that for the data point in the unoccupied state, we refer to the latest LDA calculation of Bi2212[30] and assume the same renormalization factor for both the unoccupied and occupied states. The returned fitted parameters are:

| $\mu$ | $t$ | $t'$ | $t''$ | $t^{(3)}$ | $t^{(4)}$ |
|---|---|---|---|---|---|
| 0.2130 eV | 0.1944 eV | -0.0338 eV | 0.0305 eV | 0.0028 eV | -0.0060 eV |

# References:

1. Bednorz, J. G. & Müller, K. A. Possible high $T_c$ superconductivity in the Ba-La-Cu-O system. *Z. Phys. B* **64**, 189–193 (1986).
2. Shen, Z. X. & Dessau, D. S. Electronic structure and photoemission studies of late transition-metal oxides—Mott insulators and high-temperature superconductors. *Phys. Rep.* **253**, 1–162 (1995).
3. Damascelli, A., Hussain, Z. & Shen, Z. X. Angle-resolved photoemission studies of the cuprate superconductors. *Rev. Mod. Phys.* **75**, 473–541 (2003).
4. Ronning, F. *et al*. Anomalous high-energy dispersion in angle-resolved photoemission spectra from the insulating cuprate $Ca_2CuO_2Cl_2$. *Phys. Rev. B* **71**, 094518 (2005).
5. Graf, J. *et al*. Universal High Energy Anomaly in the Angle-Resolved Photoemission Spectra of High Temperature Superconductors: Possible Evidence of Spinon and Holon Branches. *Phys. Rev. Lett.* **98**, 067004 (2007).
6. Xie, B. P. *et al*. High-Energy Scale Revival and Giant Kink in the Dispersion of a Cuprate Superconductor. *Phys. Rev. Lett.* **98**, 147001 (2007).
7. Valla, T. *et al*. High-Energy Kink Observed in the Electron Dispersion of High-Temperature Cuprate Superconductors. *Phys. Rev. Lett.* **98**, 167003 (2007).
8. Meevasana, W. *et al*. Hierarchy of multiple many-body interaction scales in high-temperature superconductors. *Phys. Rev. B* **75**, 174506 (2007).
9. Chang, J. *et al*. When low- and high-energy electronic responses meet in cuprate superconductors. *Phys. Rev. B* **75**, 224508 (2007).
10. Inosov, D. S. *et al*. Momentum and Energy Dependence of the Anomalous High-Energy Dispersion in the Electronic Structure of High Temperature

Superconductors. *Phys. Rev. Lett.* **99**, 237002 (2007).
11. Zhang, W. *et al*. High Energy Dispersion Relations for the High Temperature $Bi_2Sr_2CaCu_2O_8$ Superconductor from Laser-Based Angle-Resolved Photoemission Spectroscopy. *Phys. Rev. Lett.* **101**, 017002 (2008).
12. Inosov, D. S. *et al*. Excitation energy map of high-energy dispersion anomalies in cuprates. *Phys. Rev. B* **77**, 212504 (2008).
13. Macridin, A., Jarrell, M., Maier, T. A. & Scalapino, D. J. High-Energy Kink in the Single-Particle Spectra of the Two-Dimensional Hubbard Model. *Phys. Rev. Lett.* **99**, 237001 (2007).
14. Byczuk, K. *et al*. Kinks in the dispersion of strongly correlated electrons. *Nature Phys.* **3**, 168-171 (2007).
15. Markiewicz, R. S., Sahrakorpi, S. & Bansil, A. Paramagnon-induced dispersion anomalies in the cuprates. *Phys. Rev. B* **76**, 174514 (2007).
16. Tan, F., Wan, Y. & Wang, Q. H. Theory of high-energy features in single-particle spectra of hole-doped cuprates. *Phys. Rev. B* **76**, 054505 (2007).
17. Zhu, L., Aji, V., Shekhter, A. & Varma, C. M. Universality of Single-Particle Spectra of Cuprate Superconductors. *Phys. Rev. Lett.* **100**, 057001 (2008).
18. Schwaller, P. *et al*. Structure and Fermi surface mapping of a modulation-free Pb-Bi-Sr-Ca-Cu-O high-temperature superconductor. *J. Electron Spectrosc. Relat. Phenom.* **76**, 127 (1995).
19. Bansil, A., Lindroos, M., Sahrakorpi, S. & Markiewicz, R. S. Role of site selectivity, dimensionality, and strong correlations in angle-resolved photoemission from cuprate superconductors. *New J. Phys.* **7**, 140 (2005).
20. Shirley, D. A. High-Resolution X-Ray Photoemission Spectrum of the Valence Bands of Gold. *Phys. Rev. B* **5**, 4709 (1972).
21. Dessau, D. S. *et al*. Key features in the measured band structure of $Bi_2Sr_2CaCu_2O_{8+\delta}$: Flat bands at $E_F$ and Fermi surface nesting. *Phys. Rev. Lett.* **71**, 2781 (1993).
22. Campuzano, J. C. *et al*. Electronic Spectra and Their Relation to the $(\pi,\pi)$ Collective Mode in High- $T_c$ Superconductors. *Phys. Rev. Lett.* **83**, 3709 (1999).
23. Liu, L. Z., Anderson, R. O. & Allen, J. W. Crucial role of inelastic background subtraction in identifying non-Fermi liquid behavior in existing arpes lineshape data. *J. Phys. Chem. Solids* **52**, 1473 (1991).
24. Koralek, J. D. *et al*. Laser based angle-resolved photoemission, the sudden approximation, and quasiparticle-like spectral peaks in $Bi_2Sr_2CaCu_2O_{8+\delta}$. *Phys. Rev. Lett.* **96**, 017005 (2006).
25. Anderson, P. W. Comment on ''Anomalous spectral weight transfer at the superconducting transition of $Bi_2Sr_2CaCu_2O_{8+\delta}$''. *Phys. Rev. Lett.* **67**, 660 (1991).
26. Laughlin, R. B. Evidence for Quasiparticle Decay in Photoemission from Underdoped Cuprates. *Phys. Rev. Lett.* **79**, 1726 (1997).
27. Casey, P. A., Koralek, J. D., Plumb, N. C., Dessau, D. S. & Anderson, P. W. Accurate theoretical fits to laser-excited photoemission spectra in the normal phase of high-temperature superconductors. *Nature Phys.* **4**, 210 - 212 (2008).

28. Landau, L. D. The theory of a Fermi liquid. *Sov. Phys. JETP*, **3**, 920 (1957).

29. Varma, C. M., Littlewood, P. B. & Schmitt-Rink, S. Phenomenology of the normal state of Cu-O high-temperature superconductors. *Phys. Rev. Lett.* **63**, 1996 (1989).

30. Lin, H., Sahrakorpi, S., Markiewicz, R. S. & Bansil, A. Raising Bi-O Bands above the Fermi Energy Level of Hole-Doped $Bi_2Sr_2CaCu_2O_{8+\delta}$ and Other Cuprate Superconductors. *Phys. Rev. Lett.* **96**, 097001 (2006).

## Acknowledgements:

The authors thank R. Markiewicz for helpful discussions. This work was supported by the U.S. Department of Energy under Grant No. DE-FG02–03ER46066. The Advanced Light Source is supported by the Director, Office of Science, Office of Basic Energy Sciences, of the U.S. Department of Energy under Contract No. DE-AC02–05CH11231. The SSRL is operated by the Office of Basic Energy Sciences, of the U.S. Department of Energy under Contract No. DE-AC03-765F00515.

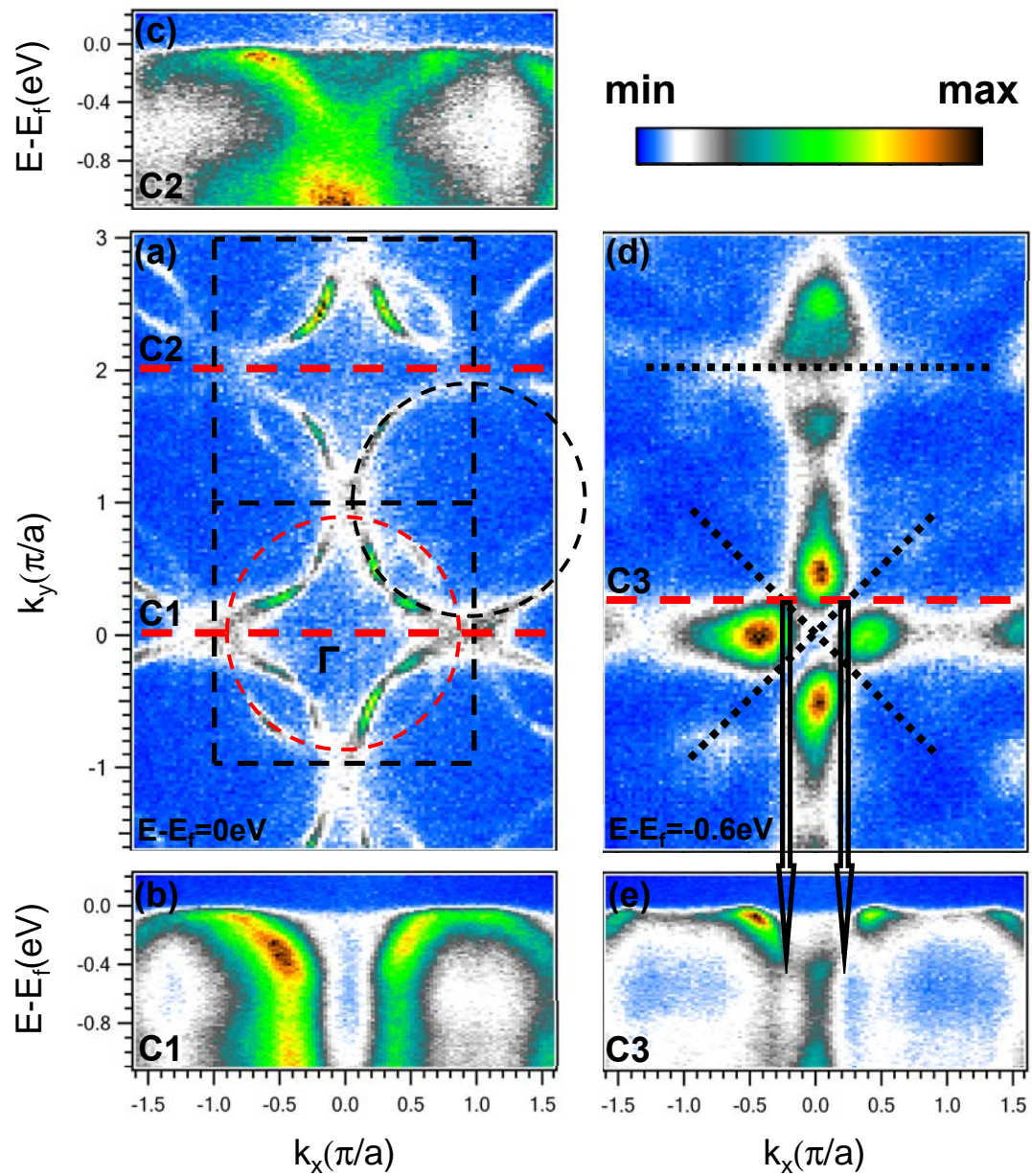

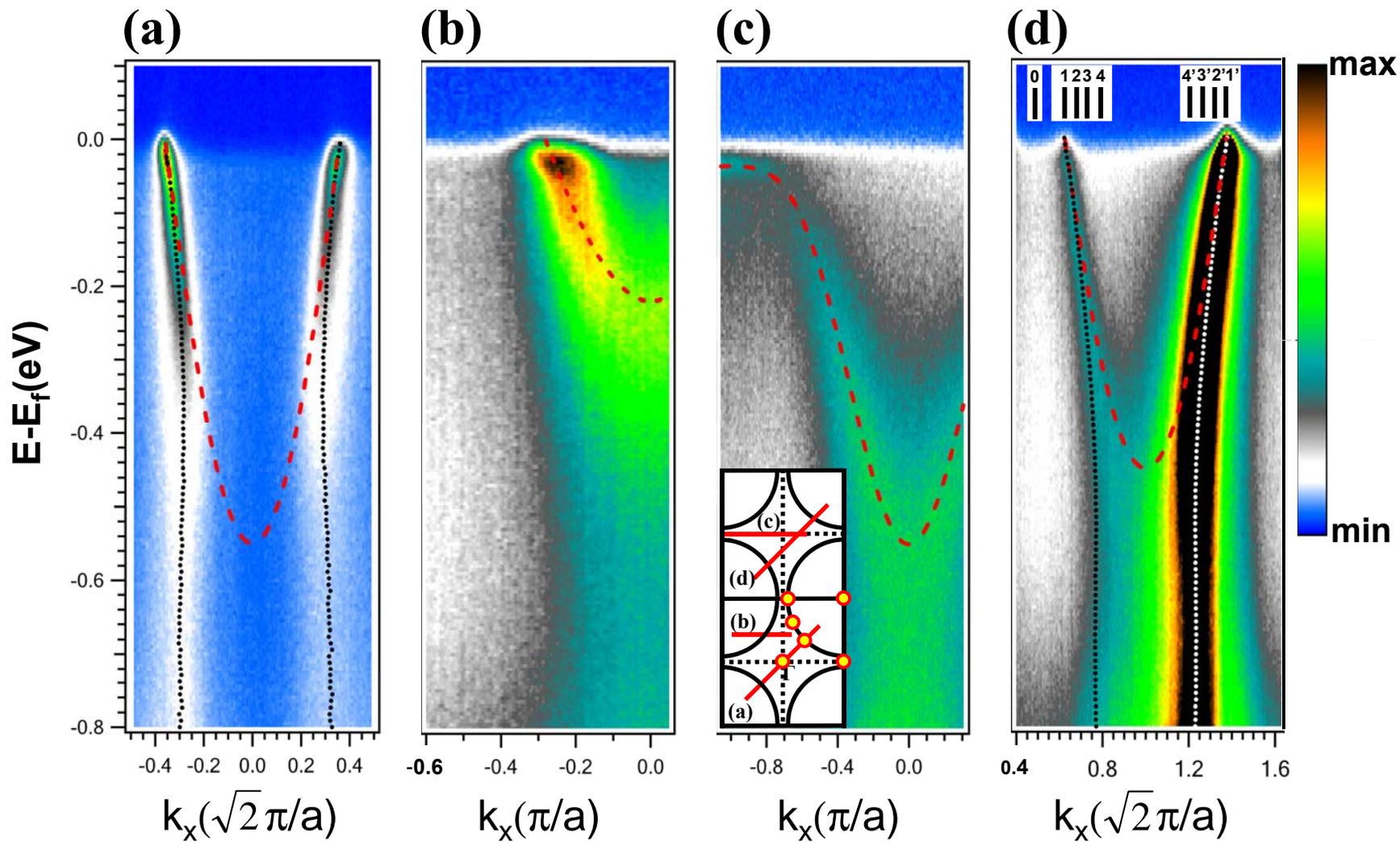

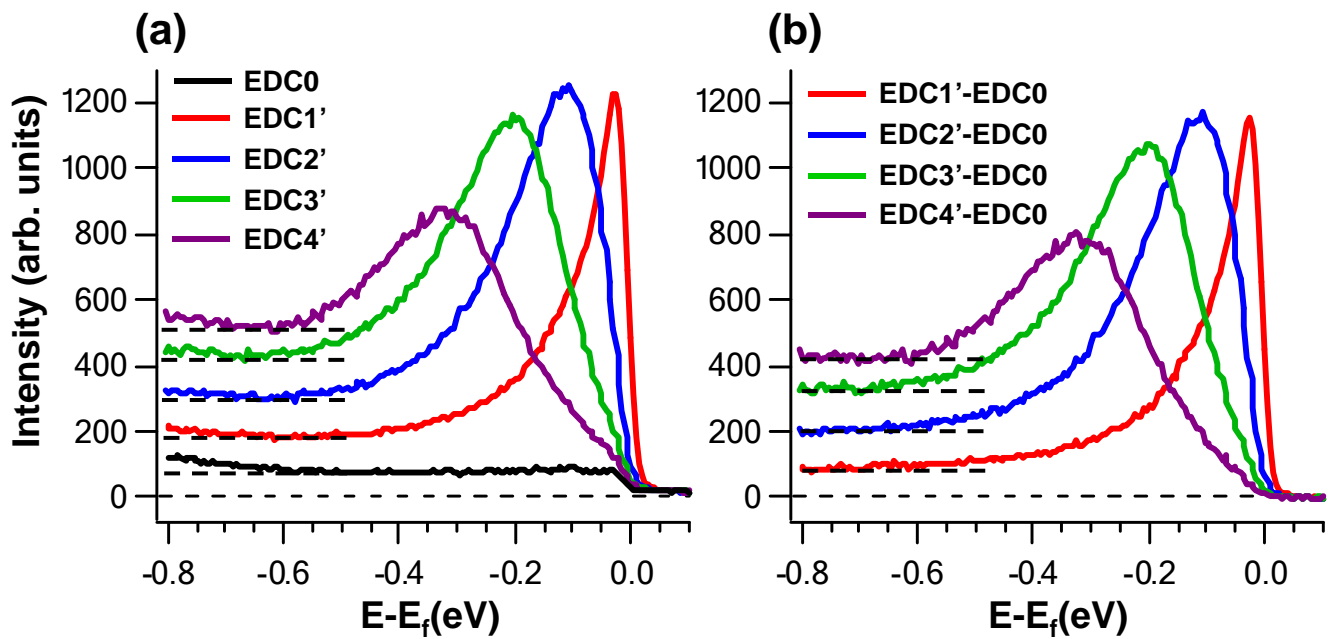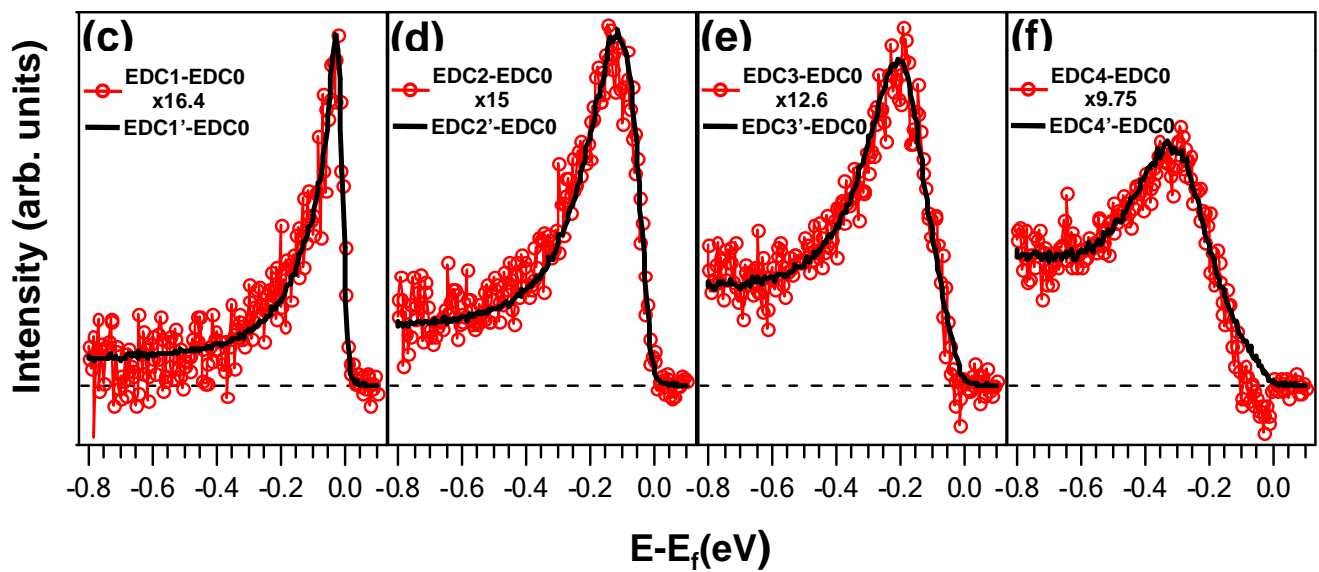

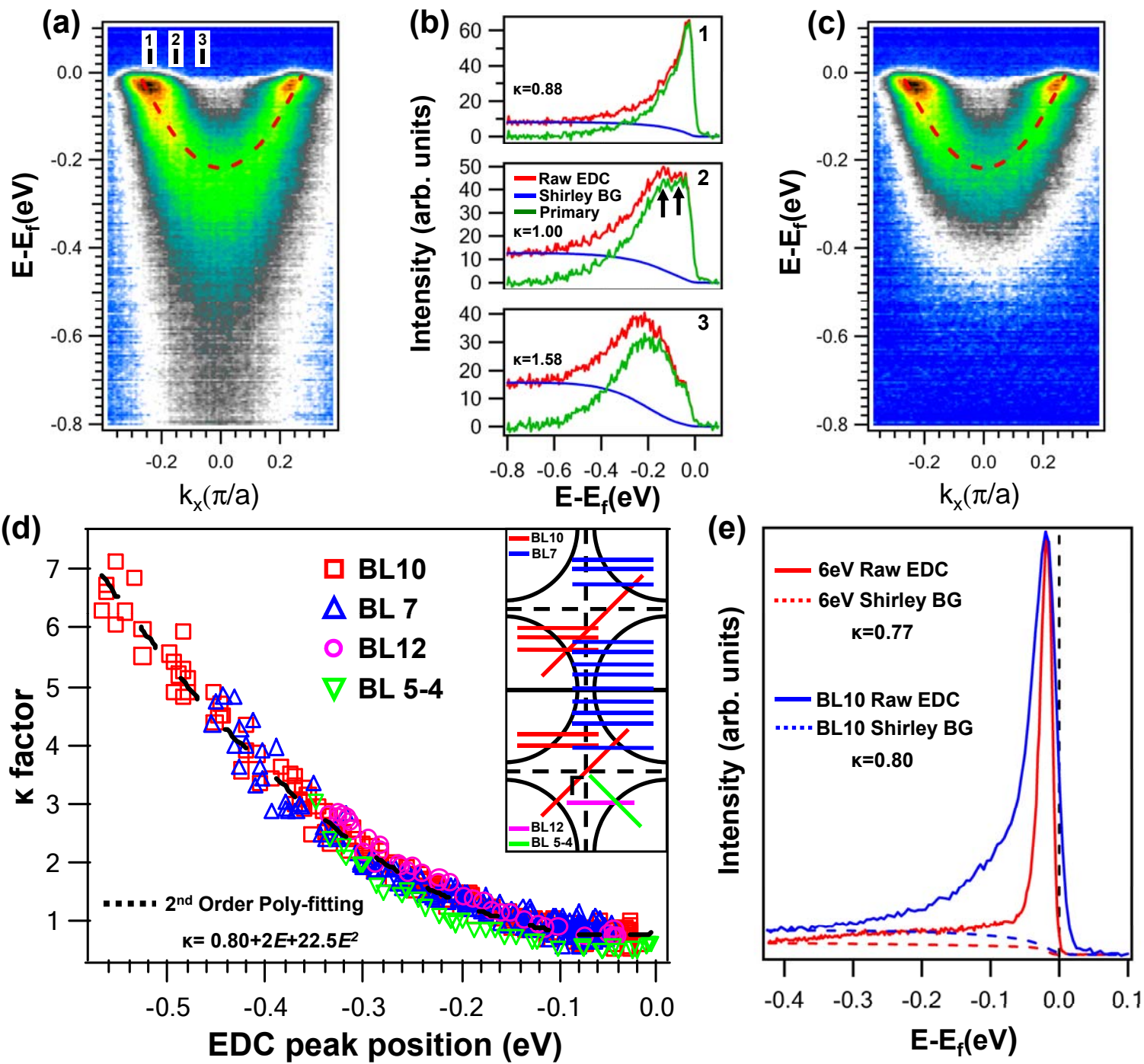